\documentclass[fleqn,usenatbib]{mnras}

\usepackage{graphicx}   
\usepackage{amsmath}    
\usepackage{amssymb}    
\usepackage{txfonts}
\usepackage{natbib,longtable,times,multicol}
\usepackage{caption}
\usepackage[caption = false]{subfig}
\usepackage{verbatim}
\usepackage{makecell}
\usepackage{epstopdf}
\usepackage{bm}

\newcommand{\bc}{\begin{center}}
\newcommand{\ec}{\end{center}}
\usepackage{color}

\title[Clustering of pseudo bulge and classical bulge galaxies ]
      {The clustering of galaxies with pseudo bulge and classical bulge in the local Universe}
\author[L. Wang, et al.]
       {Lan Wang$^{1,2}$\thanks{Contact e-mail: \href{mailto:wanglan@bao.ac.cn}{wanglan@bao.ac.cn}}, 
        Lixin Wang$^{3}$\thanks{Contact e-mail: \href{mailto:wanglixin2016@tsinghua.edu.cn}{wanglixin2016@tsinghua.edu.cn}},
        Cheng Li$^{3}$, 
        Jian Hu$^{2}$, 
        Houjun Mo$^{3,4}$, 
        Huiyuan Wang$^{5}$
        \\      
        $^1$Key Laboratory for Computational Astrophysics, National Astronomical Observatory, 
            Chinese Academy of Sciences, \\ Datun Road 20A, Beijing 100101, China\\
        $^2$National Astronomical Observatory, 
            Chinese Academy of Sciences, Datun Road 20A, Beijing 100101, China\\
		$^3$Tsinghua Center for Astrophysics and Physics Department, 
		    Tsinghua University, Beijing 100084, China\\
        $^4$Department of Astronomy, University of Massachusetts Amherst, MA 01003, USA\\
        $^5$Key Laboratory for Research in Galaxies and Cosmology, Department of Astronomy,
            University of Science and Technology of China, \\ Hefei, Anhui 230026, China}

\date{Accepted 2018 ???? ??. 
      Received 2018 ???? ??; 
      in original form 2018 ???? ??}
\pubyear{2018}

\begin{document}

\label{firstpage}

\pagerange{\pageref{firstpage}--\pageref{lastpage}} 
\maketitle

\begin{abstract}
We investigate the clustering properties and close neighbour counts for galaxies with different types of bulges and stellar masses. We select samples of ``classical'' and ``pseudo'' bulges, as well as ``bulge-less'' pure disc galaxies, based on the bulge/disc decomposition catalog of SDSS galaxies provided by \cite{simard2011}. For a given galaxy sample we estimate: the projected two-point cross-correlation function with respect to a spectroscopic reference sample, $w_p(r_p)$, and the average background-subtracted neighbour count within a projected separation using a photometric reference sample, $N_{neighbour}(<r_p)$. We compare the results with the measurements of control samples matched in color, concentration and redshift. We find that, when limited to a certain stellar mass range and matched in color and concentration, all the samples present similar clustering amplitudes and neighbour counts on scales above $\sim0.1h^{-1}$Mpc. This indicates that neither the presence of a central bulge, nor the bulge type is related to intermediate-to-large scale environments. On smaller scales, in contrast, pseudo-bulge and pure-disc galaxies similarly show strong excess in close neighbour count when compared to control galaxies, at all masses probed. For classical bulges, small-scale excess is also observed but only for $M_{stars}<10^{10}M_\odot$; at higher masses, their neighbour counts are similar to that of control galaxies at all scales. These results imply strong connections between galactic bulges and galaxy-galaxy interactions in the local Universe, although it is unclear how they are physically linked in the current theory of galaxy formation.
\end{abstract}


\section{Introduction}
\label{sec:intro}

A bulge is a commonly existing central component of a spiral galaxy that has more concentrated light and stars compared to the more extended disc \citep{sandage1961}. In fact, a prominent bulge component is observed in at least half of the bright spiral galaxies with stellar mass greater than $\sim10^9M_{\odot}$\citep{fisher2011}. Bulges can be divided into two different subcategories: classical bulges which have similar properties as elliptical galaxies, and pseudo bulges which are more like spiral galaxies, being bluer, more discy, more rotation-dominated and more active in terms of star formation\citep{kormendy2004}. In some studies, pseudo bulges have been further divided into flat ``discy pseudobulges'' and thick ``boxy/peanut bulges'', depending on the structure of the central components \citep{athanassoula2005}. In many cases, different types of bulges have been found to coexist in the same 
galaxy. The fraction of composite bulges in barred galaxies can be as high as 70 percent \citep{mendez2014}.

Three physical processes have been proposed for the formation and growth of galactic bulges. The first process is galaxy merging, during which materials of two or more galaxies condense into the center of a galaxy \citep{aguerri2001, hammer2005}.  Another process that could form bulge in a relative short timescale is the collapse and formation of bulges due to clumpy disc instability \citep{noguchi1999, elmegreen2008}. The third one is secular evolution of slow growth of bulges, due to bar-induced disc instability \citep{kormendy2004, athanassoula2005}.

In general, the current theoretical picture is that classical bulges are produced during rapid processes like merger and clumpy disc instabilities which happen more often at high redshifts. 
On the other hand, secular evolution affects in a long term the appearance and growth of pseudo bulges, and is still shaping morphologies of galaxies at the present day \citep{obreja2013, bulge2016}. Detailed studies of the formation of bulges also indicate possibilities beyond this general picture. For example, mergers as well as external accretion of gas could also lead to the formation and growth of pseudo bulges \citep{eliche2013, guedes2013, querejeta2015, sauvaget2018}, while wet major mergers
could produce a disc galaxy with both classical bulge and pseudo bulge \citep{athanassoula2016}. On the other hand, secular evolution can build up some massive bulges in the early universe, without the help of major mergers \citep{genzel2008}. 

Nevertheless, the current picture can not fully explain the observational properties of bulges. In particular, many giant galaxies have been observed to have no sign of a classical bulge, a result that is inconsistent with the hierarchical clustering cosmology which predicts the opposite, indicating that bulge formation may be somehow suppressed in mergers \citep{bulge2016}. This problem is a strong function of environment: while most of the giant galaxies in the Local Group are bulgeless or with pseudo bulges, galaxies in Virgo cluster are mostly ellipticals or with classical bulges. In order to loose the tension, minor mergers have been proposed, as an important channel particularly responsible for the formation of pseudo bulges \citep[e.g.][]{eliche2013},
without destroying the existing thin disc in galaxies \citep[e.g.][]{moster2010}. However, there has been little observational evidence in support of this picture. In fact, very few pseudo bulge galaxies show signs of tidal interactions of minor mergers \citep{kormendy2004}.

While mergers happen more often in denser environment where galaxies have more companions, merger-induced bulges might cluster more strongly than bulges formed through internal processes. In this work, we use the SDSS decomposition catalogue provided by \citet{simard2011} to select observed galaxies with different morphologies, and compare their clustering properties. We also obtain neighbour counts of these selected galaxy samples, to study their small scale environment. By examining these statistical properties,  rather than looking into galaxies for possible merger indicators \citep[e.g.,][]{lopez2016} or studying the detailed environment of bulge galaxies \citep[e.g.,][]{henkel2017,mishra2017b}, we hope to understand further the role merger and environment play in the formation of different types of galactic bulges. 

The paper is organized as follows. In section 2.1 we introduce briefly the Simard et al. decomposition catalogue, and how we select galaxies of different morphologies from it. Methods of measuring galaxy correlation functions and close neighbour counts are described in Section 2.2. In section 3 we first show clustering results of different types of galaxies in four stellar mass bins, then both clustering and close neighbour count results of matched samples that have the same distributions in redshift, color and concentration. Discussions and conclusions are presented in section 4 and 5, respectively.


\section{Data and methodology}
\label{sec:sample}

\subsection{SDSS galaxy samples}
\label{sec:simard}

In this work, we make use of the SDSS galaxy catalogue of  \citet{simard2011} to select samples of classical and pseudo bulges.
\cite{simard2011} performed a two-dimensional photometric decomposition of bulge and disc components, with the point spread function being convolved, for over a million galaxies in the SDSS Data Release 7 \citep[DR7;][]{abazajian2009}. Three different fitting models were applied to each galaxy: a pure Sersic profile, a two-component model consisting of a Sersic profile for the bulge and an exponential profile for the disc, and the same two-component model except that the Sersic index $n_b$ is fixed to $n_b=4$.  Detailed fitting methods can be found in \citet{simard2011}, as well as some examples of fitting results (their Fig.5 and Fig.6).
To quantify the robustness of the model fitting, two F-test probabilities are provided: 1) $P_{pS}$: the probability that the two-component model with a free Sersic index $n_b$ is not required compared to a pure Sersic model, and 2) $P_{n4}$: the probability that the two-component model with a free $n_b$ is not required compared to the two-component model with a fixed index of $n_b=4$.
\citet{simard2011} showed that, a system truly consisting of both bulge and disc components can be robustly identified by requiring $P_{pS} \leq 0.32$. In their catalog ~26\% galaxies meet this requirement. However, a Sersic index $n_b$ is robustly determined only for 9\% of the bulge+disc systems, for which $P_{n4} \leq 0.32$ is required. Nevertheless, given the large size of the parent catalog, we're able to select samples of galaxies with well classified bulge types, which are substantially large for our purpose.

\begin{figure*}
\bc
\hspace{-1.4cm}
\resizebox{15cm}{!}{\includegraphics{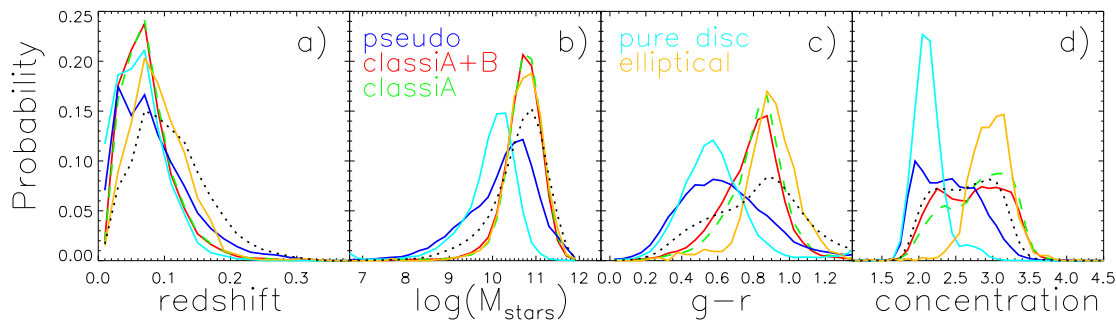}}\\%
\caption{
Statistics of the three selected samples (color lines as labeled).
From left to right, panels show distributions of: a) redshift; b) stellar mass;
c) petrosian magnitude g-r color; and d) concentration.
Black dotted lines are the distributions of galaxy properties in
the whole sample, without any selection of morphology. 
}
\label{fig:sampledis}
\ec
\end{figure*}

We start with all the galaxies in the \cite{simard2011} catalog with spectroscopically measured redshifts available from SDSS/DR7. This forms our parent sample, consisting of 692292 galaxies, from which we construct a number of subsamples according to the presence of a bulge and/or the bulge type. First, we select two subsamples of galaxies, with either a ``pseudo bulge'' or a ``classic bulge'' in their center. For this purpose, following \cite{simard2011}, we first select galaxies with $P_{pS} \leq 0.32$ and $P_{n4} \leq 0.32$, which are expected to be truly a two-component system consisting of an exponential disc plus a bulge with a free index parameter of $n_b$. We further divide these galaxies into two sub-catagories, with a bulge type of ``pseudo'' or ``classic'', according to the empirical divider of $n_b = 2$ \citep{kormendy2004}. 
Pseudo bulges are selected to have $n_b < 2$, and classical bulges are the ones with $n_b > 3$. Galaxies with $n_b = [2,3]$ are not considered in order to minimize uncertainty in the classification. 
Studies of nearby bulge-disc galaxies showed that around 90 per cent of classical (pseudo) bulges have $n_b>2$ ($n_b<2$) \citep[Figure 1.3 of ][]{fisher2016}. With less than $\sim 10$ per cent contamination, we expect the statistical results like clustering and neighbor count as we study in this work not to be strongly biased. On the other hand, although a combination of several criteria may be useful to select out the most typical bulge samples, it would decrease the numbers of samples by a large fraction. 

In addition, we select the galaxies with $P_{n4} \geq 0.68$ as the third subsample. These galaxies also have two components, a disc plus a bulge, but the bulge component is better fitted when the Sersic index is fixed to $n_b=4$. Therefore these galaxies are also in the class of classical bulge, but are not included in the classical bulge sample with $P_{n4} \leq 0.32$ and a free $n_b$ as defined above.

For comparison, we also select bulge-less pure disc galaxies and elliptical galaxies from the same parent catalog. Galaxies with $P_{pS} \geq 0.68$ are the ones better fitted by a pure Sersic model than a bulge + disc model, and are considered as either a pure disc galaxy or an elliptical galaxy. We then divide the two types according to best-fit Sersic index, requiring a pure disc to have $n_g<2$ and an elliptical to have $n_g>3$. Galaxies with an intermediate $n_g$ are not considered for the same reason as above. 

The allowed range of Sersic index $n$ ($n_b$ for pseudo bulge and classical bulge, and $n_g$ for pure disc and elliptical) in the \citet{simard2011} sample is $0.5<n<8$. Following \citet{mishra2017a}, we exclude galaxies with $n\geq 7.95$ in our samples, since high values of Sersic indices may be associated with fitting problems \citep{meert2015}. \citet{mishra2017a} also exclude galaxies that have error of $n$ greater than $+ 1 \sigma$ of error distribution, and exclude galaxies that host a bar when selecting pseudo bulges. We have checked that when removing galaxies with large errors in $n$, the results remain similar as when including them. We therefore do not exclude these galaxies to get better statistics. The distributions of error in $n$ as a function of $n$ value in our samples are presented in Appendix A. We also check the effect of galaxies having bars on the determination of Sersic index in Appendix A. 

Apart from the commonly used Sersic index criterion \citep{kormendy2004,fisher2008} to select pseudo bulge and classical bulge as we do in this work, galaxies with different types of bulges also have statistically difference on some other properties \citep[a complete list of properties that define bulge types can be found in ][]{kormendy2016}. In some studies to define galaxies with bulges \citep[e.g.][]{vaghmare2015, mishra2017a}, positions on the Kormendy diagram \citep{kormendy1977} and velocity dispersion are also used to select the purest pseudo bulge sample. 
 In our work, while we aim to get better statistics for both classical and pseudo bulge galaxies, our pseudo bulges sample selected based only on Sersic index is not necessarily made of the purest pseudo bulges as in \citet{vaghmare2015} and \citet{mishra2017a}. 
In Appendix B, we discuss and check these additional criteria for galaxies in our selected samples, and test the effect of including them on our results. 

We check the robustness of our identification of pure disc and elliptical samples by cross-matching our samples with the morphology catalogues of \citet{nair2010} and \citet{nicholas2018}. In the matched galaxies, 97.6 (74.0) per cent of our pure discs (ellipticals) have probability of being spiral(elliptical) greater than 0.5 in \citet{nicholas2018}. 85.9 (86.7) per cent of our pure discs (ellipticals) have types later (earlier) than S0 in \citet{nair2010}. Our type determinations of pure discs and ellipticals are largely consistent with these samples of more detailed morphology classification. 

In summary, we have selected five samples: three samples for galaxies with different types of bulges, one for pure-disc galaxies and one for elliptical galaxies. Table~\ref{tab:samples} lists the name, the selection criteria and the size of all the samples. Since the number of galaxies in Classical B is relatively small for calculating correlation functions, in the following section we do not analyse Classical B individually, but rather work on a combined sample of Classical A and Classical B, which is referred to
as {\tt Classical A+B}.

\begin{table}
    \caption{ Samples of galaxies with different morphologies and bulge types selected from the \citet{simard2011} catalog.}
    \label{tab:samples}
    \begin{tabular}{llr}
    \hline
    Sample & Selection Criteria & Size \\
    \hline
    Pseudo bulge        & $P_{pS} \leq 0.32$, $P_{n4} \leq 0.32$, $n_b < 2$ & 19,813 \\[2pt]
    Classical bulge A   & $P_{pS} \leq 0.32$, $P_{n4} \leq 0.32$, $3<n_b<7.95$  & 15,906 \\[2pt]
    Classical bulge B   & $P_{pS} \leq 0.32$, $P_{n4} \geq 0.68$, $n_b = 4$ & 7,219  \\[2pt]
    Pure disc           & $P_{pS} \geq 0.68$, $n_g < 2$ & 11,102 \\[2pt]
    Elliptical          & $P_{pS} \geq 0.68$, $3<n_g < 7.95$ &  990 \\[2pt]
    \hline
    \end{tabular}
\end{table}

Fig.~\ref{fig:sampledis} displays the the distributions of redshift, stellar mass, $g-r$ color and the concentration parameter for our samples. Stellar masses are taken from the MPA-JHU SDSS catalogue \footnote{The catalogue is publicly available at: $http://www.sdss.org/dr12/spectro/galaxy{\_}mpajhu/$}. The concentration parameter is defined as the ratio of two radii, enclosing 90 and 50 percent of the galaxy light in the r band \citep[see][]{stoughton2002}. Compared with the whole sample without selection of galaxy morphologies as shown in black dotted lines, the galaxies selected in our samples have in general lower redshifts, which can be understood from the fact that closer galaxies are better determined in morphology type. pure disc galaxies peak at low stellar mass, blue color and low concentration, while ellipticals peak on the opposite ends. As expected, pseudo bulge galaxies are similar to pure disc galaxies in distributions of stellar mass and color, while galaxies with classical bulges are more like ellipticals in these properties. The distributions of bulge galaxies in concentration, however, are much wider than that of pure disc/elliptical galaxies, showing a relative flat trend in a large range of concentration. 

Comparing pseudo bulge galaxies with classical bulge galaxies, we find the former to be generally less massive, bluer, and less concentrated. Different bulge types show quite a large overlap in color and concentration space, indicating that these properties can not be simply used as criteria to distinguish classical bulges from pseudo bulges. 

\begin{figure*}
\bc
\hspace{-0.4cm}
\resizebox{17cm}{!}{\includegraphics{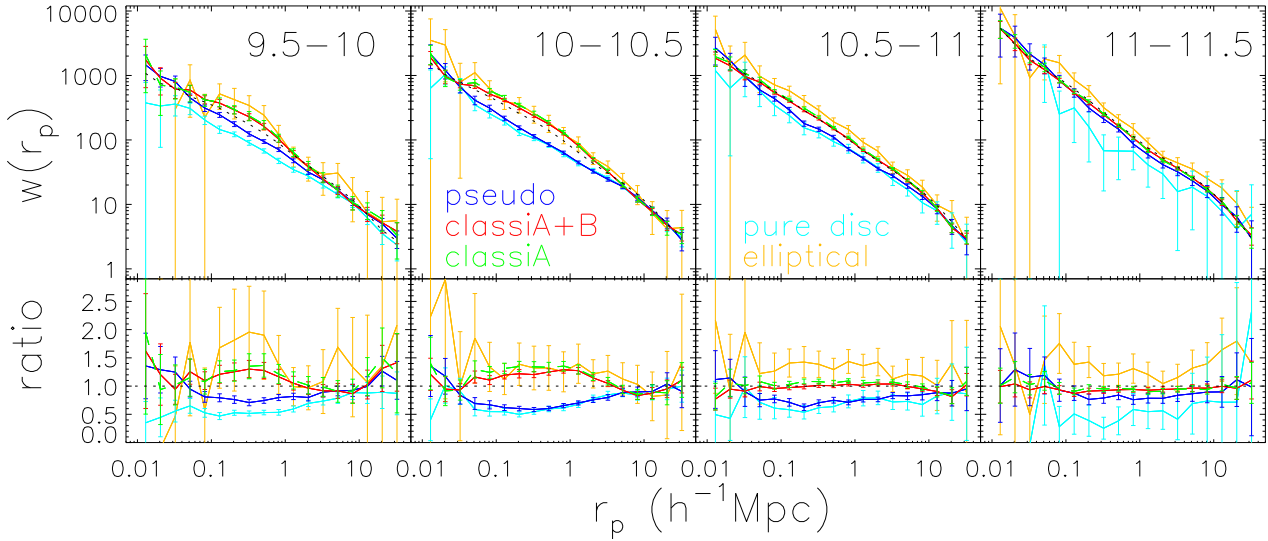}}\\%
\caption{
Upper panels: clustering of galaxies in different stellar mass bins, indicated 
in range of $log(M_{stars}/M_{\odot})$ in black. Blue, red, green, cyan and gold lines are
results of pseudo, classical A, classical A+B, pure disc and elliptical samples. 
Black dots are results of the whole Simard et al. sample, 
matched with the same stellar mass criterion. Bottom panels: for given stellar mass bin,
the ratio between 2PCCF of galaxies in selected morphology samples and that 
of galaxies in the whole sample. 
}
\label{fig:CFMstars}
\ec
\end{figure*}

\subsection{Methodology}

\subsubsection{Two-point cross-correlation functions}
\label{sec:CF}

In this work we use two-point cross-correlation function (2PCCF) to quantify the clustering of our galaxies. Below we describe briefly the methodology of measuring the 2PCCF. Detailed description can be found in \citet{li2006,licheng2006b,li2012}. 

For a given galaxy sample {\tt Q}, the 2PCCF is measured with respect to a reference galaxy sample {\tt G}, which is a magnitude-limited galaxy catalog selected from the SDSS/DR7-based NYU-VAGC catalog {\tt dr72}, consisting of about half a million galaxies with $r$-band Petrosian apparent magnitude limited to $r < 17.6$, $r$-band absolute magnitude in the range of $-24 < M^{0.1}_r < -16$, and spectroscopically-measured redshift in the range of $0.01 < z < 0.5$. Both $r$ and $M^{0.1}_r$ are corrected for Galactic extinction, and $M^{0.1}_r$ is also corrected for evolution and K-corrected to its value at $z = 0.1$. The NYU-VAGC catalog is publicly available \footnote{NYU-VAGC catalog can be found at:  $http://sdss.physics.nyu.edu/vagc/$}, and described in detail in \cite{blanton2015vagc} and \cite{blanton2017kcorrect}. We have constructed a random sample {\tt R}, which has the same selection effects as, but 10 times larger than the reference sample, following the method described in \citet{li2006}. 

A redshift-space 2PCCF, $\xi^{(s)}(r_p,\pi)$, between Samples Q and G, is first estimated by:
\begin{equation}
\xi^{(s)}(r_p,\pi) = \frac{N_{R}}{N_{G}}\frac{QG(r_{p},\pi)}{QR(r_{p},\pi)}-1,
\end{equation}
where $r_{p}$ and $\pi$ are the pair separations perpendicular and parallel to the line of sight; $N_{R}$ and $ N_{G}$ are the number of galaxies in the random sample ({\tt R}) and in the reference sample {\tt G};  $QG(r_{p},\pi)$ and  ${QR(r_{p},\pi)}$
are the counts of cross pairs between samples {\tt Q} and {\tt G}, and between samples {\tt Q} and {\tt R}. The projected 2PCCF, $w_p(r_p)$, is then obtained by integrating $\xi^{(s)}(r_p,\pi)$ over $\pi$:
\begin{equation}
w_{p}(r_{p}) = \int_{-\infty}^{+\infty}\xi(r_{p},\pi)d\pi = \sum\xi(r_{p},\pi_{i})\Delta\pi_{i}. 
\end{equation}
The summation for computing $w_p(r_p)$ runs from $\pi_1=-40h^{-1}$Mpc to $\pi_{80}=40h^{-1}$Mpc, with $\Delta\pi_{i} = 1 h^{-1} Mpc$. The effect of fiber collisions is corrected using the method detailed in \citet{li2006}. Errors on all the $w_p(r_p)$ measurements are estimated using the bootstrap resampling technique \citep{barrow1984}.

\begin{figure}
\bc
\hspace{-0.4cm}
\resizebox{8cm}{!}{\includegraphics{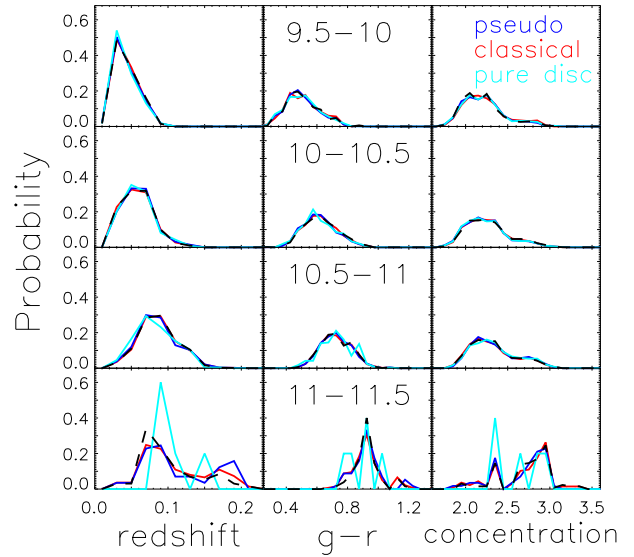}}\\
\caption{
For four stellar mass bins as indicated in range of $log(M_{stars}/M_{\odot})$ in black,
panels in each row show at the given stellar mass bin, the
distributions of redshift, color and concentration of galaxies in matched samples. 
Black dashed lines are distributions of the corresponding control samples.
}
\label{fig:samedis}
\ec
\end{figure}

\subsubsection{Close neighbour counts}
\label{sec:NC}

In addition to 2PCCFs, we count the number of companion galaxies in the vicinity of the galaxies in our samples, using a photometric reference sample which is also constructed from the NYU-VAGC version {\tt dr72} by selecting galaxies with $r$-band apparent magnitude down to $r<21$. The photometric reference sample includes about 2.5 million galaxies over the same sky area of the galaxy samples being studied. We make a statistical correction for the effect of chance projections on the neighbour counts, by subtracting the average count at the same scale around a large number of randomly placed galaxies. When compared to 2PCCFs, close neighbour counts are not affected by fiber collisions on small scales, and can include much fainter companion galaxies thanks to the photometric sample which is much deeper than the spectroscopic sample, thus probing the effect of close companions over a broader range of mass ratios. Detailed description of the method of estimating close neighbour counts, as well as tests and example applications can be found in \cite{li2008a,li2008b}.

\section{results}
\label{sec:results}

\subsection{Joint dependence of clustering on galaxy mass and morphology}
\label{sec:CFmass}

Galaxy clustering depends on stellar mass, with more massive galaxies being clustered more strongly \citep{li2006}. Therefore, we first divide galaxies in each of our samples into four intervals of stellar mass, and estimate the 2PCCF $w_p(r_p)$ for each of them. For each of the stellar mass interval, Fig.~\ref{fig:CFMstars} compares the $w_p(r_p)$ measurements for galaxies of different morphological types. The bottom panels in the same figure display the ratios of the $w_p(r_p)$ measured from our samples relative to the $w_p(r_p)$ measured for all the galaxies in the same stellar mass range as selected from the whole sample of \cite{simard2011}.  

\begin{figure*}
\bc
\hspace{-0.4cm}
\resizebox{17cm}{!}{\includegraphics{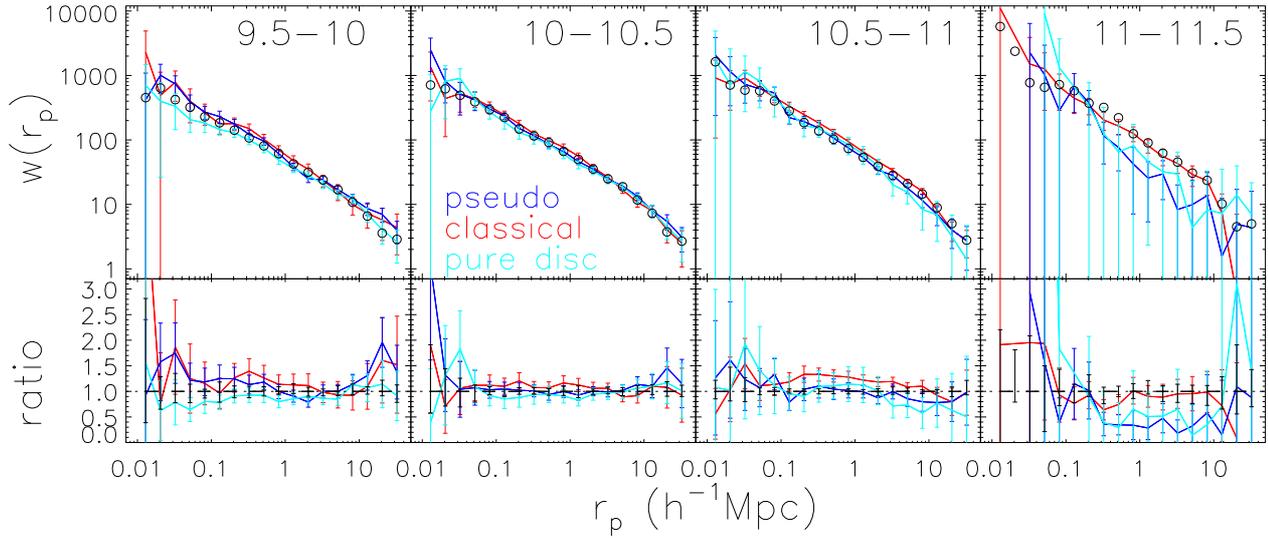}}\\
\caption{
Clustering of galaxies in four stellar mass bins, for galaxy samples matched with
the same distributions of redshift, color and concentration in given stellar mass bin. 
The stellar mass intervals
are indicated in range of $log(M_{stars}/M_{\odot})$ in black. For each stellar mass interval,
upper panel gives 2PCCF clustering results of pseudo bulge galaxies (blue line),
classical A+B bulge galaxies (red lines), and pure disc galaxies (cyan lines). Black circles 
are the median result of the 20 control samples constructed. 
Error bars on matched galaxy samples are bootstrap errors. The corresponding lower
panel shows the ratios between 2PCCF of selected galaxy samples and that of the
control sample. Black error bars indicate the 68 percentile distributions of 
the 20 control samples constructed.
}
\label{fig:CF}
\ec
\end{figure*}

\begin{figure*}
\bc
\hspace{-0.4cm}
\resizebox{17cm}{!}{\includegraphics{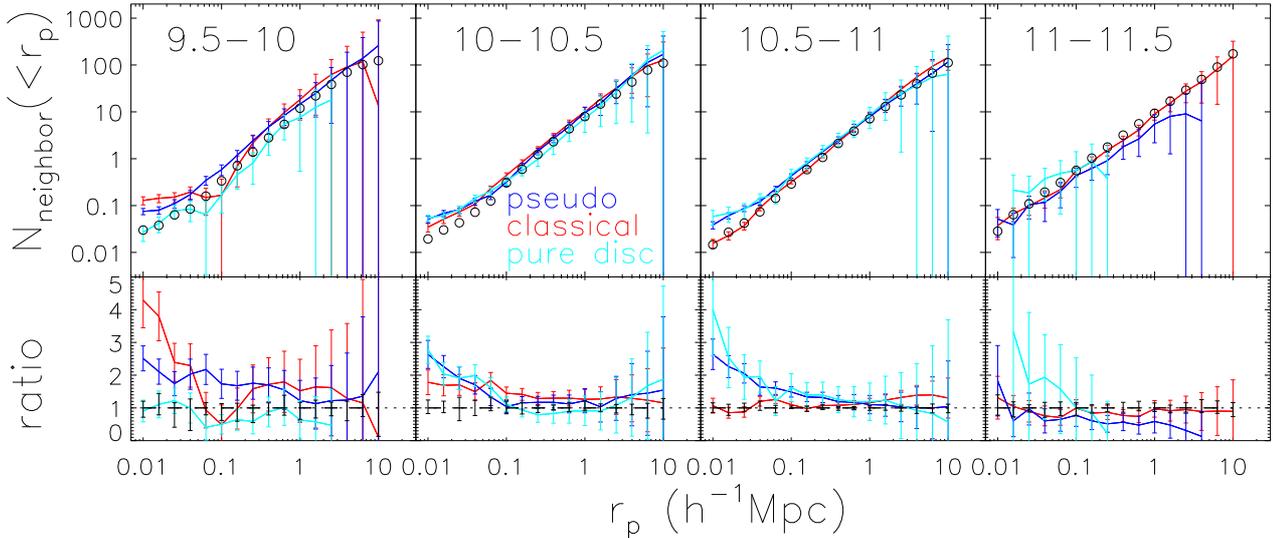}}\\
\caption{
Neighbour counts of galaxies in four stellar mass bins indicated in range of 
$log(M_{stars}/M_{\odot})$ in black, for galaxy samples matched with the same 
distributions of redshift, color and concentration in each panel.
Upper panels are the average counts of galaxies in the photometric sample to an 
r-band limiting magnitude of $r_{lim}$ = 20 within a given projected radius 
$r_p$ from the selected galaxies. Blue/red solid line gives result of 
pseudo/classical A+B bulge samples, and cyan solid line is for pure disc samples,
with bootstrap errors shown. Black circles give the median results of the 20 
control samples in each stellar mass bin. 
Lower panels are the ratios between the neighbour counts of the morphology 
selected samples to the median results of the control samples, with black error bars 
indicating the 68 percentile distributions of the 20 control samples. 
}
\label{fig:neighbourcount}
\ec
\end{figure*}

Overall, the figure shows that all the samples appear to cluster similarly at both smallest scales ($r_p\lesssim 0.1 h^{-1}$kpc) and scales larger than a few Mpc. At intermediate scales, different samples show different clustering behaviors. First, galaxies with classical bulges present stronger clustering than those with pseudo bulges, with larger differences at lower masses, and the difference becomes indistinguishable when stellar mass exceeds $10^{11} M_{\odot}$. Second, although the error bars are large for the pure disc and elliptical galaxies due to small sample sizes, there is still an obvious trend that pure disc galaxies show weakest clustering and ellipticals appear to cluster most strongly, in a given stellar mass range. It is interesting that galaxies with pseudo bulges seem to show very similar clustering behaviors to the pure disc galaxies of similar mass, while galaxies with classical bulges are more like ellipticals in terms of mass dependence of clustering. Finally, when compared to the whole sample, the pseudo bulges are clustered more weakly at all masses except the highest mass bin, while the clustering amplitude of classical bulges  is comparable to that of the whole sample at all masses except the lowest masses at which the classical bulges are more strongly clustered.  

\begin{figure}
\bc
\hspace{-0.4cm}
\resizebox{8cm}{!}{\includegraphics{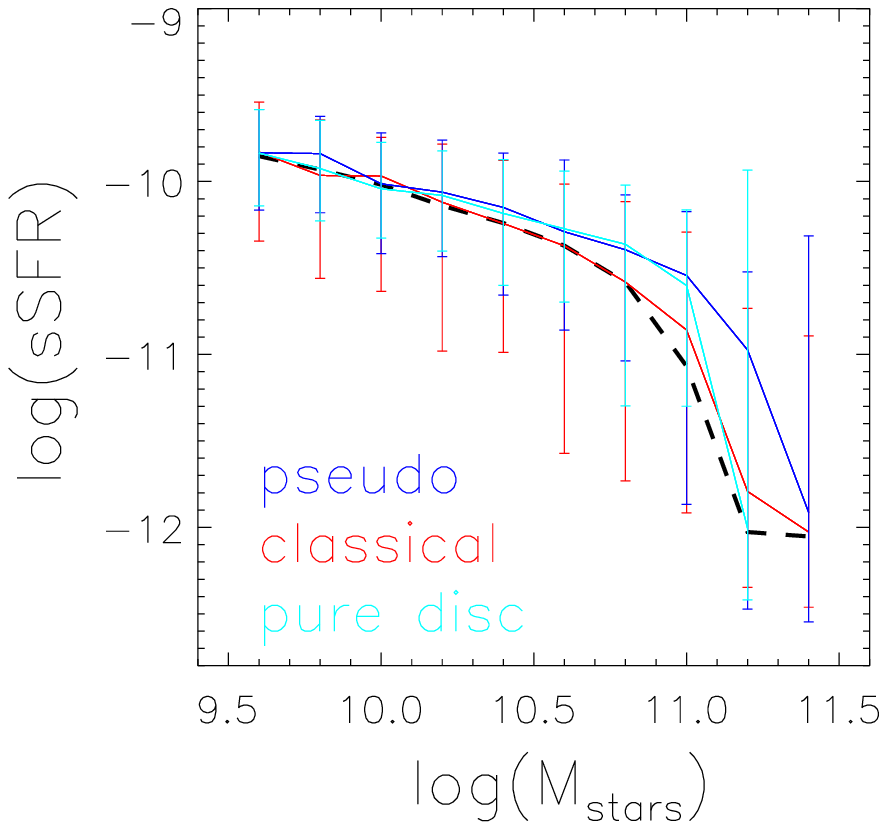}}\\%
\caption{
 sSFR of the matched bulge samples and pure disc sample (solid lines are median value, and error bars give 68 ~per~cent distribution) as a function of galaxy stellar mass. Blue, red and cyan lines are for pseudo bulge, classical bulge and pure disc samples respectively.
Median result of one corresponding control samples is shown by black dashed line.
}
\label{fig:sSFR}
\ec
\end{figure}

\begin{figure*}
\bc
\hspace{-0.4cm}
\resizebox{17cm}{!}{\includegraphics{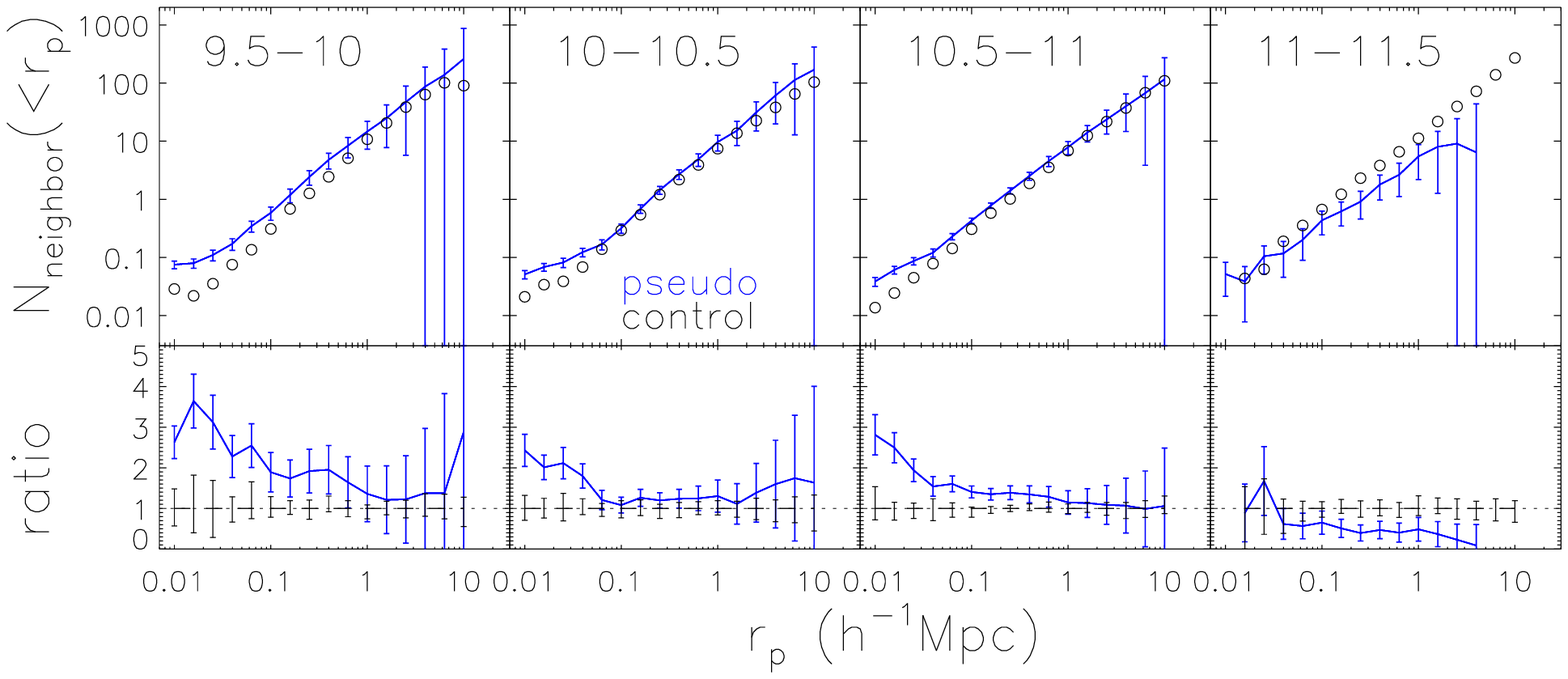}}\\  
\caption{
Similar as Fig.~\ref{fig:neighbourcount}, but for control samples matched to 
have the same distributions of redshift, color, concentration and sSFR, with the pseudo 
bulge samples as shown in Fig.~\ref{fig:neighbourcount}.
}
\label{fig:NC_sSFR}
\ec
\end{figure*}

Previous studies have well established that, at given stellar mass, galaxy clustering depends also on other properties, such as color and structural parameters, with stronger clustering for galaxies with redder colors and more centrally concentrated light distributions \citep[e.g.][]{li2006}. In order to exclude the effect of such residual dependence, for a certain stellar mass bin, we have trimmed the galaxy samples in such a way that they have the same distributions in $g-r$ color, concentration parameter and redshift. From what follows we will not consider the elliptical galaxy sample which is too small after trimmed to allow any meaningful statistics. As mentioned above, we will combine the two bulge samples ``classical A'' and ``classical B'' into a single sample in order for better statistics. We note that we have repeated all the following analyses using the sample of ``classical bulge A'' alone, finding pretty much the same results as what we find from the merged sample. Fig.~\ref{fig:samedis} shows the distributions of redshift, $g-r$ and concentration for the three samples: ``pseudo bulge'', ``classical bulge'' and ``pure disc'', after they are trimmed. At given stellar mass, the three samples are matched very well in all these parameters. 

For each given stellar mass bin, we have constructed a set of 20 control samples to be compared with the three matched samples. The control samples are randomly selected from the reference sample {\tt G} as mentioned in sec.~\ref{sec:CF}, each required to have the same distributions in redshift, color and concentration as the galaxy samples. The number of galaxies included in each control sample is the same as the largest of the matched samples in the stellar mass bin considered. The distributions of redshift, color and concentration for the control samples are plotted as black dashed lines in Fig.~\ref{fig:samedis}. 

\begin{figure*}
\bc
\hspace{-0.4cm}
\resizebox{17cm}{!}{\includegraphics{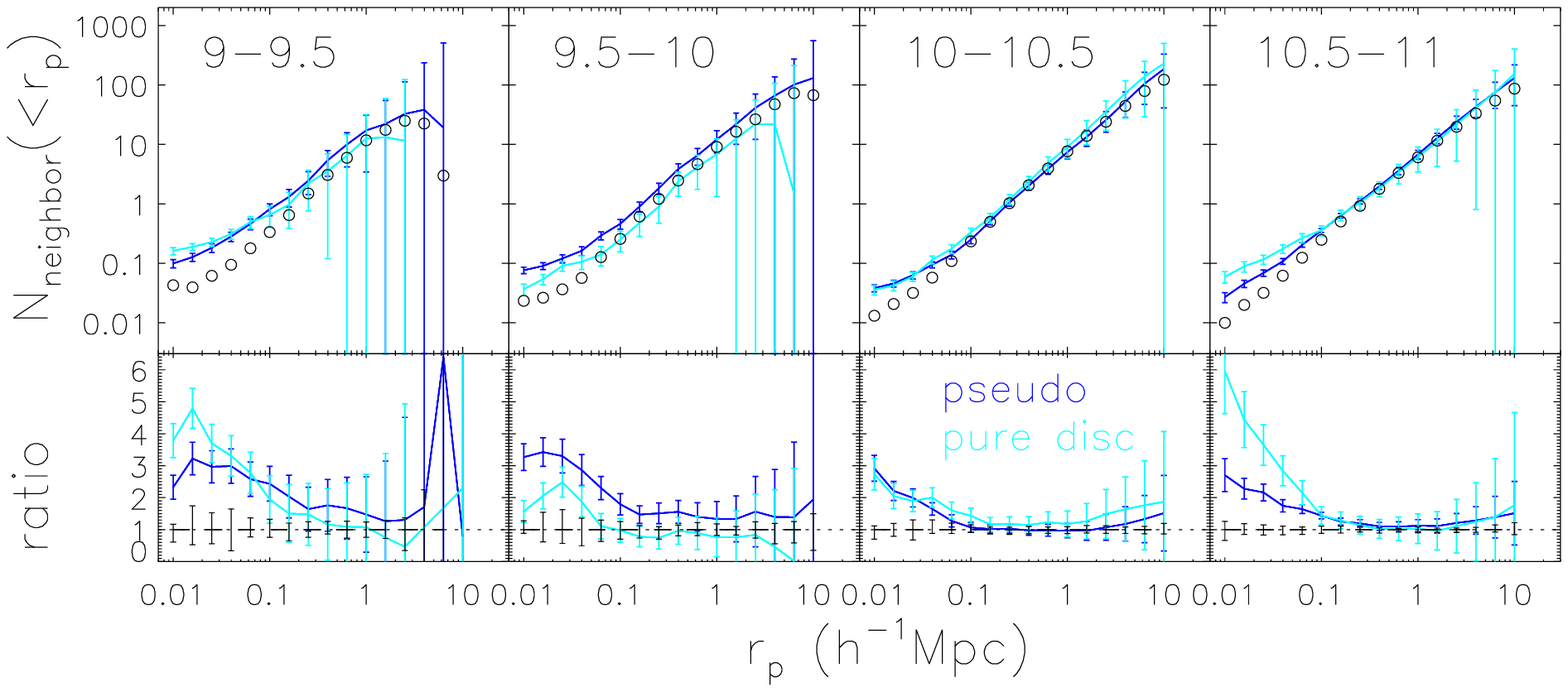}}\\
\caption{
Similar as Fig.~\ref{fig:neighbourcount}, but for pseudo bulge and pure disc galaxies 
matched with the same distributions
of redshift, color and concentration, while control samples are also constructed to have
the same distributions. Stellar mass bins are in the range of $10^{9-11}M_{\odot}$.
}
\label{fig:NC_pseu_disk}
\ec
\end{figure*}

\begin{figure*}
\bc
\hspace{-0.4cm}
\resizebox{15cm}{!}{\includegraphics{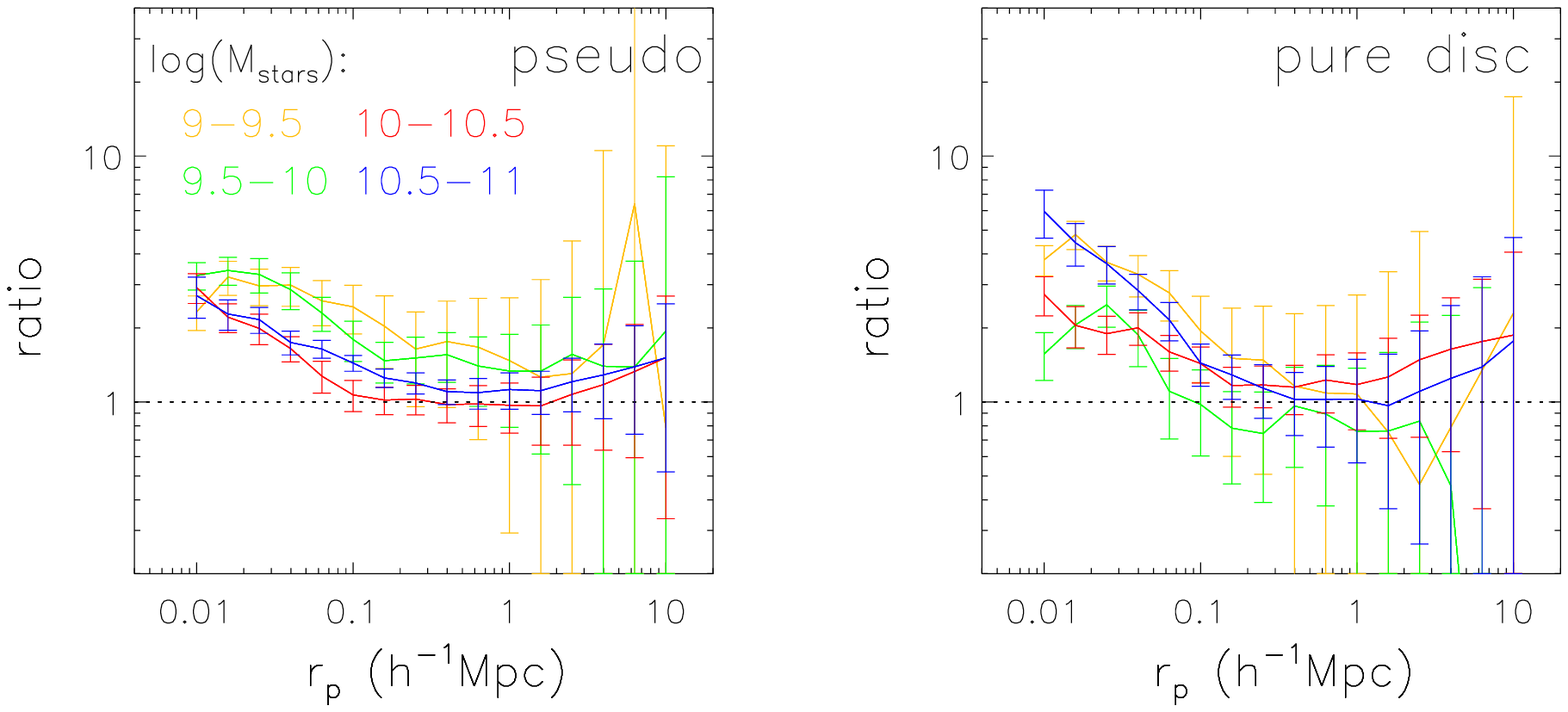}}\\
\caption{
The ratio between neighbour counts of the matched pseudo bulge (left panel) and pure disc (right panel) galaxy samples and that of the corresponding control sample, for four stellar mass bins 
indicated by different color. Error bars show bootstrap errors of the matched galaxy samples.
}
\label{fig:NC_Mstar}
\ec
\end{figure*}

The clustering results of the matched galaxy samples and control samples in different stellar mass bins are shown in Fig.~\ref{fig:CF}. For control samples, the median of the 20 control samples is shown as black circles. Ratios of the 2PCCF of the three matched samples relative to the corresponding control samples (black circles) are presented in the lower panels. 
In general, the differences between different morphology samples and control samples are small, mostly within error bars, except for the most massive bin. Therefore, the significant differences between classical bulge galaxies and pseudo bulge/pure disc galaxies as seen in Fig.~\ref{fig:CFMstars} should be largely attributed to their different distributions of redshift, color and concentration. For the most massive bin, there seems to have a trend that both pure disc galaxies and pseudo bulge galaxies cluster less strongly than classical bulge galaxies and the control samples. We would not overemphasize this trend, however, given the large errors in the mass bin.

We note that, at smallest scales with $r_p\lesssim 0.1 h^{-1}$Mpc and in some stellar mass bins, both the pure disc galaxies and the pseudo bulge galaxies tend to present a sharply increasing $w_p(r_p)$ when compared to the control samples. However, considering both the small number of pair counts at these small scales and the possible residual effect of SDSS fiber collisions, we are unable to make any firm conclusions based on the $w_p(r_p)$ measurements at these scales. In the next subsection, we will focus on these small scales by estimating close neighbour counts. To summarize the analyses presented in this subsection, we have found no significant differences in clustering at scales above $0.1 h^{-1}$Mpc for all the samples considered in our study.

\subsection{Probing the small-scale environment by counting close neighbours}
\label{sec:density}

We have estimated the background-subtracted close neighbour count as a function of projected radius, $N_{neighbour}(<r_p)$, in the vicinity of both the samples of different morphology types and the corresponding control samples, using the photometric reference sample down to a given limiting magnitude of $r_{lim}$ as described in \S~\ref{sec:NC}. Fig.~\ref{fig:neighbourcount} displays the results obtained with $r_{lim}=20$, separately for the different stellar mass bins. In a given stellar mass bin, all the samples including the samples of pseudo bulge, classical bulge and pure disc galaxies, as well as the control samples, are closely matched in $g-r$ color, concentration and redshift as in the previous subsection. First of all, we see that when the stellar mass is limited to a certain range, the neighbour counts at scales larger than $\sim0.1 h^{-1}$Mpc for all the samples agree within errors, which are too large to deduce any trend. This is well consistent with the 2PCCF comparison results as presented above.

On small scales with $r_p$ less than a few  $\times100 h^{-1}$kpc, the neighbour counts reveal a number of interesting results, which could not be seen above from the 2PCCFs. First, galaxies with classical bulges have similar numbers of neighbours to the control samples, and this is true for all the scales probed and at stellar masses above $10^{10} M_{\odot}$. In the lowest mass bin, the classic bulges have significantly higher counts at $r_p<0.1 h^{-1}$Mpc than all the other samples. Second, in contrast to classic bulges, pseudo bulges in the highest mass bin show weak enhancement in the smallest-scale neighbour count compared to control samples, but the ratio of the neighbour counts increases to $\sim 2-3$ for all the other mass bins. Finally, for pure disc galaxies, a positive correlation with stellar mass is clearly seen, in the sense that the neighbour count ratio is flat at unity at the lowest masses but increases with increasing mass, reaching a factor of $\sim 4$ at the smallest scales in the highest two mass bins. 

It is striking to see the highly enhanced neighbour counts with respect to the control samples, which are observed at $r_p<0.1 h^{-1}$Mpc for most of the samples of both pseudo-bulge galaxies and pure disc galaxies, although the $N_{neighbour}$ ratio 
depends on mass. At $r_p=0.1 h^{-1}$Mpc, the average count of neighbours around both types of galaxies is around 0.3 for stellar mass in the range $10^{10-10.5}M_\odot$, and more than 0.4 for stellar mass in the range $10^{10.5-11}M_\odot$, 
which means, more than 40\% of the galaxies in these samples have a companion within 100  $h^{-1}$kpc. This fraction is unexpectedly high, which is comparable to or even higher than the neighbour counts found ($\sim 0.4$) for the most strongly star-forming galaxies (with specific SFR of $\log(sSFR) >$ -8.8) in the SDSS, as measured by \citet{li2008a} using the same photometric reference sample and the same methodology (see their Fig.11). 

 Since both pseudo bulge galaxies and pure disc galaxies may have strong star formation, their large neighbour counts at small scales as seen in Fig.~\ref{fig:neighbourcount} (also later in Fig.~\ref{fig:NC_pseu_disk}) may be partly (if not entirely) a result of their high star formation rates. This is not the case, however, as can be seen from Fig.~\ref{fig:sSFR} which compares the distributions of our samples in the plane of specific SFR (sSFR=SFR/$M_{stars}$) versus stellar mass. The pseudo-bulge galaxies and pure disc galaxies in our sample have  very similar median sSFR, and are similar to that of the classical bulges and control sample galaxies at stellar masses less than $\sim 10^{10.5}M_{\odot}$. At stellar mass greater than $\sim 10^{10.5}M_{\odot}$, pseudo bulges and pure disc samples have a bit higher median sSFR than the control samples, but far from as high as the most star forming galaxies in \citet{li2008a} that have similar enhancement in close neighbour count.
For this figure we have taken the SFR measurements from the MPA-JHU SDSS catalogue (see section \ref{sec:sample}).

We have further examined the possible contribution of the $N_{neighbour}$-sSFR connection to our results by additionally matching the pseudo-bulge samples with the control samples in sSFR. The neighbour counts and the ratios to the control samples are shown in Fig.~\ref{fig:NC_sSFR}. Although the overall amplitudes of the $N_{neighbour}$ ratio become more noisy, the general trends and our conclusions remain unchanged. Apparently the small-scale enhancement in the neighbour counts as found in our samples and the similar enhancement as previously found by
\citet{li2008a} in strongly star-forming galaxies are not the same effect. 

\subsection{Close neighbour enhancement in pseudo bulge and pure disc samples}
\label{sec:test}

 In the stellar mass range of $10^{10-11}M_\odot$, pseudo bulge and pure disc galaxies show similar excess of neighbour counts with respect to the control samples. At lowest and highest mass bins, the error bars become relatively large, due to the small sizes of the matched galaxy samples when all the samples of pseudo bulge, classical bulge and pure disc in a given stellar mass bin are required to be closely matched in various properties. In order to see the difference between pseudo bulge and pure disc galaxy samples with better statistics, we have repeated the analysis in Fig.~\ref{fig:neighbourcount}, but matching only pseudo bulge and pure disc samples to have similar color, concentration and redshift. The results in the stellar mass range of $10^{9-11}M_\odot$ are shown in Fig.~\ref{fig:NC_pseu_disk}. The errors of the neighbour counts and their ratios to the control samples are reduced as expected, particularly for pure disc galaxy samples, and qualitatively the results remain the same as what we see from Fig.~\ref{fig:neighbourcount} --- the neighbour count amplitudes at scales smaller than $\sim0.1 h^{-1}$Mpc are significantly enhanced when compared to control galaxies of similar mass, for both pseudo bulges and pure disc galaxies and at all stellar masses. 
 
 Fig.~\ref{fig:NC_pseu_disk} shows that, quantitatively, the $N_{neighbour}$ ratio of the pseudo bulges presents a clear anti-correlation with stellar mass, with the small-scale $N_{neighbour}$ ratio decreasing from $\sim3$ at the lowest masses to $\sim2$ at the highest masses. For pure disc galaxies, the $N_{neighbour}$ ratio at the small scales depends on stellar mass in a non-monotonic manner --- the small-scale enhancement in the neighbour count is lowest at $M_{stars}=10^{9.5-10}M_\odot$, and increases at both higher and lower masses. 

The non-monotonic mass dependence of pure disc galaxies can be seen more clearly in Fig.~\ref{fig:NC_Mstar} (see the right panel), where we show the pure disc galaxy-to-control ratio as a function of $r_p$ for all the stellar mass bins in a single panel. The $N_{neighbour}$ ratios are consistent at unity within errors on scales larger than $\sim 0.1 h^{-1}$Mpc, and increases significantly at smaller scales, reaching a value of 5 or 6 at the smallest scales in both the lowest mass bin ($10^9<M_{stars}<10^{9.5}M_\odot$) and the highest mass bin ($10^{10.5}<M_{stars}<10^{11}M_\odot$). At the intermediate masses, the ratio is at levels of 2-3. In the left-hand panel of the same figure, the results are plotted in the same way for the samples of pseudo-bulge galaxies. The $N_{neighbour}$ ratios also go beyond unity at small scales with $r_p<0.1 h^{-1}$Mpc, but they are highest at the lowest masses and decrease with increasing mass monotonically at fixed scale.

\begin{figure*}
\bc
\hspace{-0.4cm}
\resizebox{15cm}{!}{\includegraphics{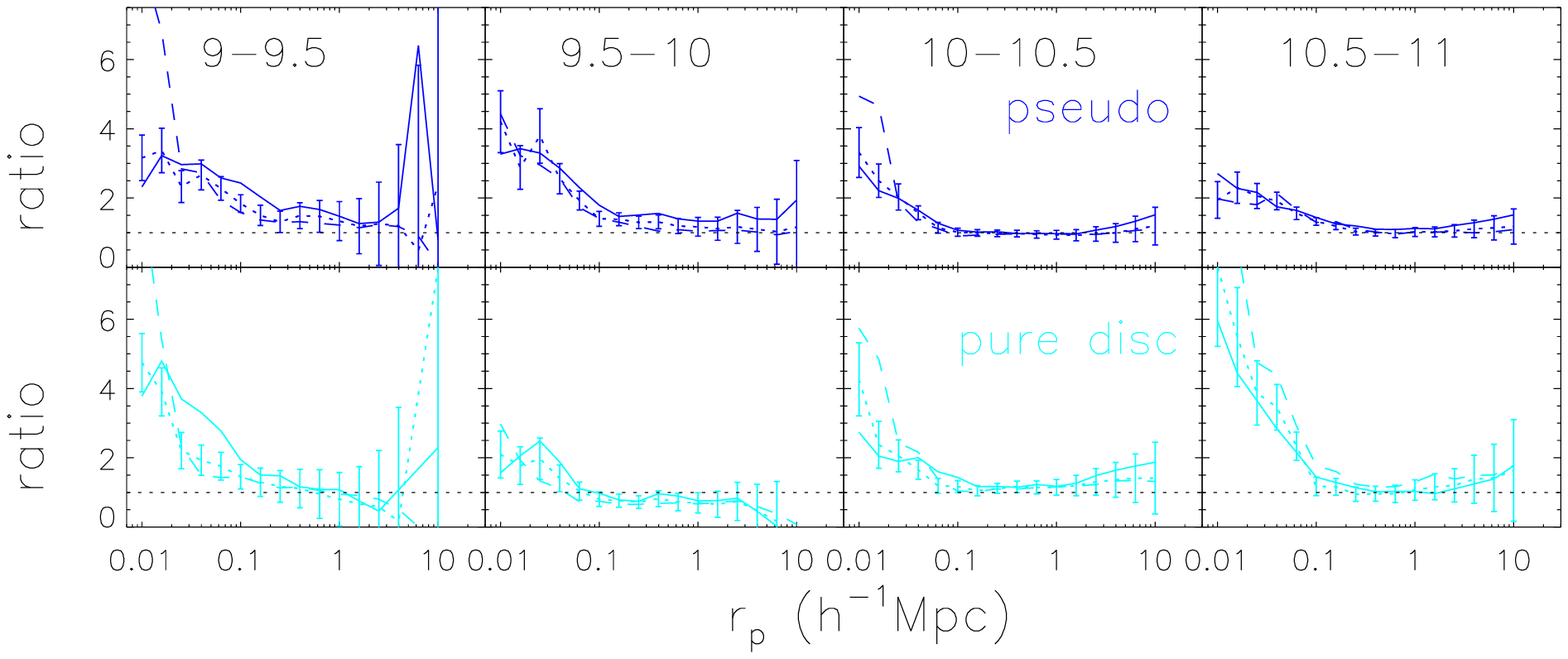}}\\
\caption{
The ratio between neighbour count of matched pseudo bulge (upper panels) and pure disc (bottom panels) 
galaxy samples and that of the corresponding control sample, for four stellar mass bins. 
Solid, dotted and dashed lines corresponds to the reults of 
r-band limiting magnitude of $r_{lim}$ = 20, 19, and 18 respectively. 
Error bars on the dotted lines show bootstrap errors for results of $r_{lim}$ = 19.
}
\label{fig:NC_rlim}
\ec
\end{figure*}

The neighbour counts we have obtained so far are based on the photometric sample down to a $r$-band limiting magnitude of $r_{lim}$ = 20. In Fig.~\ref{fig:NC_rlim} we examine the dependence of neighbour counts on $r_{lim}$, showing the $N_{neighbour}$ ratios between the matched galaxy samples and the control samples for three different magnitude limits: $r_{lim}=$20, 19, and 18.
Pseudo bulge and pure disc samples are the same as shown in Fig.~\ref{fig:NC_pseu_disk} and Fig.~\ref{fig:NC_Mstar}. Each panel compares the results of the three limiting magnitudes but for a given stellar mass bin, and results for the pseudo-bulge samples and those for the pure disc galaxy samples are shown in upper and lower panels, respectively. In general, for a given stellar mass bin, the $N_{neighbour}$ ratio depends very weakly on $r_{lim}$. This indicates that the excess of the neighbour counts on small scales is dominated by relatively bright companions with $r<18$, while fainter neighbours contribute little. 

\section{Discussion}
\label{sec:discussion}

\begin{figure*}
\bc
\hspace{-0.4cm}
\resizebox{17cm}{!}{\includegraphics{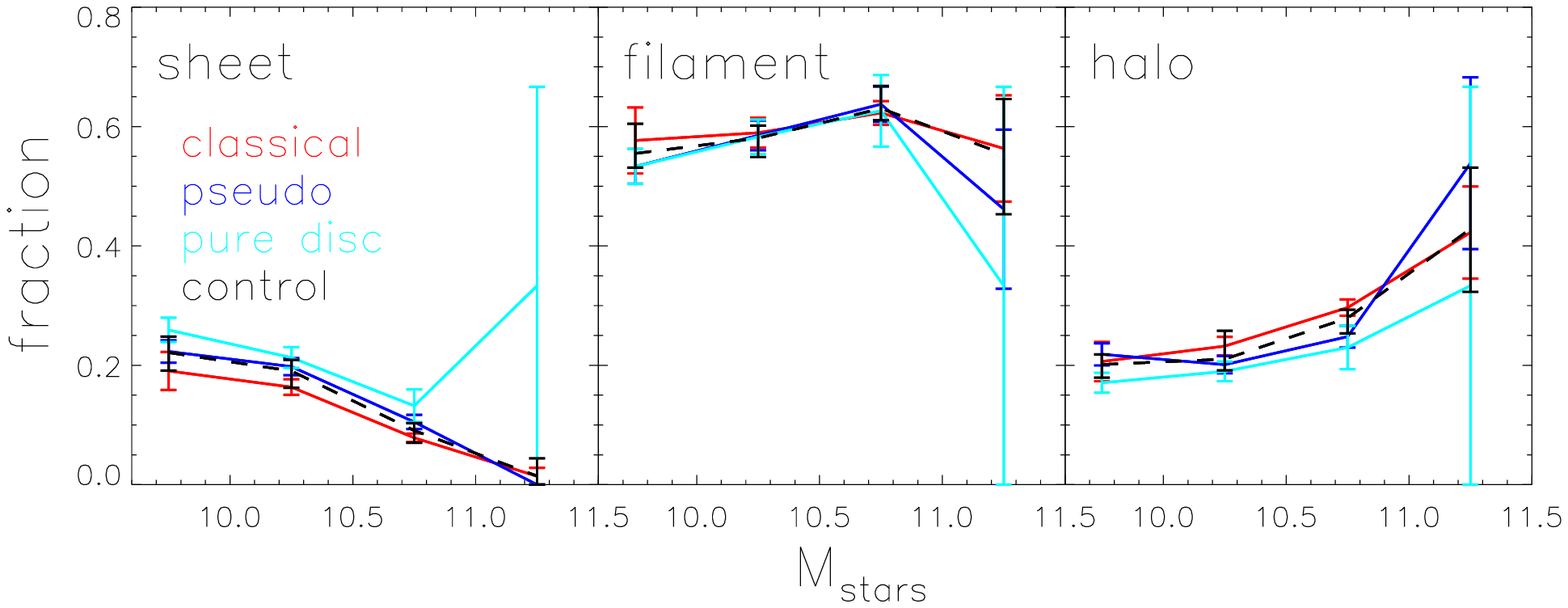}}\\
\caption{
Fraction of galaxies in different large scale structures (left panel: sheet; middle
panel: filament; right panel: halo) as a function of galaxy stellar mass,
for galaxy samples and control samples matched with the same 
distributions of redshift, color and concentration in each stellar mass bin.
The error bars of the bulge samples are Poisson errors, where for the control samples, 
dashed lines with error bars give the median value and the variation range of the 
20 control samples constructed for given stellar mass bin.
}
\label{fig:LSS}
\ec
\end{figure*}

\subsection{bulge and large-scale environment}
\label{sec:large}

On scales larger than a few $\times100 h^{-1}$kpc we see no difference in both the two-point cross-correlations and the neighbour counts for all of our samples, when they are matched to have similar mass, color and concentration parameter. This implies that the presence of a central bulge and the bulge type in a galaxy are not affected by environmental effects occurring at intermediate-to-large scales. This result is well consistent with previous studies of galaxy clustering and environment. For instance, \cite{li2006} estimated the two-point auto-correlation function of SDSS galaxies as a function of their stellar mass, optical color and galaxy structure for spatial scales above $0.1 h^{-1}$Mpc, and found the clustering amplitude at given scale shows obvious dependence on color but no dependence on concentration and surface mass density when stellar mass is limited to a narrow range. Studies of the local environment of SDSS galaxies have led to the same conclusion: the environment of a galaxy does not relate to its overall structure once the stellar population age and stellar mass are fixed \citep[see a review by][and reference therein]{blanton2009araa}.


For pure-disc galaxies without an obvious bulge component, some of the recent studies have suggested that this class of galaxies are preferentially found in low-density regions \citep[e.g.][]{kautsch2009}, and that a subset of them with a very thin disc tend to avoid filamentary structure on large scales \citep{bizyaev2017}. We have examined the latter effect using our galaxy samples and the large-scale structure type provided by \citet{Wanghuiyuan2016}. These authors obtained the initial density field for the local Universe based on the distribution of galaxy groups in the SDSS volume, and reconstructed the three-dimensional density field of the local Universe, as well as its formation history, by running a high-resolution $N$-body constrained simulation. The density field reconstruction provides not only the local density averaged over 1-3 Mpc scale for each real galaxy in the SDSS volume, but also the type of the large-scale structure (LSS) in which the galaxy resides. The large-scale structures in the density field are classified into four different types: void, sheet, filament, and halo, determined from the density field in a dynamical way following \citet{hahn2007}. The density field is limited to local galaxies with redshifts less than $z\sim0.12$, and includes more than 70\% of our galaxies.

Fig.~\ref{fig:LSS} displays the fraction of the galaxies in our sample associated with sheet, filament or halo, as a function of stellar mass. Results for galaxy samples of classic bulge, pseudo bulge and pure disc are plotted with different colors. We don't consider the LSS type of void in this analysis because of too small a fraction of galaxies falling in this class (less than 4 per cent in all samples). In general, we see very similar fractions at given stellar mass and LSS type. The majority of the galaxies are found in filaments, with a fraction of $\sim60\%$ that is little dependent on stellar mass. About 20\% of the galaxies in the lowest mass bin are found in sheets, and the fraction decreases at higher masses, reaching zero at the highest masses. The rest $\sim20\%$ of the lowest mass bin are found in halos, but the fraction increases with increasing mass, reaching about 50\% at the highest masses. These trends are all well expected, and the similarity between the samples of different morphologies and the control samples suggests again that the presence and type of galactic bulges have little to do with large-scale environments. We note that the pure disc galaxy sample seem to have a slightly higher fraction of the sheet type and correspondingly lower fraction of the halo type, while the bulge samples tend to show a slightly lower fraction of sheet but higher fraction of halo types. However, given the error bars, these subtle differences should not be overemphasized here, and more work with larger samples is needed in order to better understand these differences. With the current data, we find no obvious evidence for any type of our galaxies to either prefer or avoid the filament structure. 


\subsection{bulge and small-scale environment}
\label{sec:small}

Different from what we've seen on large scales, on scales less than $\sim 0.1$Mpc we have observed significant excess in the close neighbour counts around our galaxies when compared to control galaxies of similar mass, color and concentration. The strength of the excess depends on both morphology/bulge type and stellar mass. For classical bulges, the small-scale excess is obviously seen only at low mass ($M_{stars}=10^{9.5-10}M_\odot$), where the ratio between the classical bulge sample and the control sample reaches $\sim5$ at the smallest scales ($r_p\sim10 h^{-1}$kpc, see Fig.~\ref{fig:neighbourcount}). If we believed that classical bulges form by major mergers as mentioned in section \ref{sec:intro}, our result may be suggesting that classical bulges in massive galaxies with $M_{stars}>10^{10}M_\odot$  are post-merger remnants, thus showing no excess in close neighbour count compared to control galaxies, and that less massive galaxies with $M_{stars}<10^{10}M_\odot$ might still be forming bulges. However, we note that our sample of classical bulges in the lowest mass bin is quite small, resulting in large errors in the neighbour counting as can be seen form Fig.~\ref{fig:neighbourcount}. The highly excess of close neighbour counts around classical bulges in low-mass galaxies as reported here need to be double checked in next works with larger samples. If it is proved by future larger samples, our result means that low mass classical bulges, unlike the more massive ones, prefer to live in relative denser local regions and may form from a different mechanism, such as interaction-induced clumpy disc instability that happens more often and form classical bulges at high redshifts.

The most striking result of our work is that both pseudo-bulge galaxies and pure-disc galaxies show significant enhancement in close neighbour counts when compared to control galaxies of similar mass, color and concentration. The result holds even when the distribution of specific star formation rate is additionally matched when constructing the control sample. In some cases, the close neighbour count enhancement is even stronger than the enhancement in the same quantity previously measured for the most-strongly star-forming galaxies in SDSS  \citep{li2008a,li2008b}. Furthermore, we found that the neighbour count enhancement doesn't change much when we include fainter and fainter neighbour galaxies in the counting, which is done by increasing the limiting magnitude of the photometric reference sample. This indicates that the neighbour count enhancement is dominantly contributed by relatively bright neighbours. Our result implies strong connections between pseudo bulges, pure-disc galaxies and galaxy-galaxy interactions, although it is not quite clear how they are physically linked with each other. 

While we are focusing on the statistics of relative large galaxy samples of different bulge types, some works analyse the properties of galaxies hosting classical and pseudo bulges based on more detailed identification of smaller samples of galaxies \citep{fernandez2014}, especially for S0 and spiral galaxies \citep{mishra2017a,mishra2018}. \citet{mishra2017b} showed that the existence of different bulge types depends on environment, consistent with our findings that pseudo bulge and classical bulge have different neighbour counts. \citet{mishra2017b} and \citet{barway2017} showed that the formation of pseudo bulges seems to be independent of environment, possibly on the same line of our results that pseudo bulges and pure discs have similar excess of close neighbour counts and hence similar small scale environment.

Recent studies of bulge formation/growth have been mostly focusing on high-z galaxies, where galaxy-galaxy interactions appear to play less important roles than previously expected. For instance, \cite{Tadaki-17} found a large fraction of the extended rotating discs at $z\sim2$ to be associated with extremely compact center, dusty and strongly star-forming, which can rapidly build up a central bulge in a few $\times10^{8}$yr. Therefore, the authors suggested that bulges are commonly formed in high-z discs by internal processes, not requiring major mergers. However, the situation is quite different when one goes to lower redshifts. In another recent study, \cite{Sachdeva-17} studied bulges in bright disc galaxies at $z\lesssim1$ using data from both the GOODS-South ($0.4<z<1$) and SDSS ($0.02<z<0.05$) samples, concluding that clump migration and secular processes alone cannot account for the bulge growth since $z\sim1$, and that accretion and minor mergers would be required to explain their data. Our result is apparently along the same line with their work.

\section{Conclusion}
\label{sec:conclusion}

In this work we have investigated the clustering and close neighbour counts for galaxies with different types of galactic bulges and stellar masses. For this purpose, we have selected samples of galaxies with ``classic'' or ``pseudo'' bulges, as well as samples of ``bulge-less'' pure disc galaxies, using the photometric catalog of \cite{simard2011} who performed a careful bulge-disc decomposition on the optical image of a large sample of galaxies in the SDSS. For a given galaxy sample, we have estimated the projected two-point cross-correlation function $w_p(r_p)$ with respect to a reference sample consisting of about half a million spectroscopically observed galaxies, and the average background-subtracted neighbour count within a projected separation $N_{neighbour}(<r_p)$ using a photometric reference sample down to $r-$band limiting magnitude of $r_{lim}=20$. In order to isolate out the known correlations between the local environment and galaxy properties such as stellar mass, color and structural parameters, we have divided the galaxies into narrow ranges of stellar mass and closely matched the samples in a given mass range, so that the samples of different morphologies and bulge types have similar distributions in redshift, $g-r$ and concentration parameter $R_{90}/R_{50}$. 

Our main conclusions can be summarized as follows:
\begin{itemize}
    \item When limited to a certain stellar mass range and closely matched in redshift, color and concentration, all the samples are found to present similar clustering amplitudes and average neighbour counts on scales larger than $\sim0.1 h^{-1}$Mpc. This indicates that neither the presence of a galactic bulge nor the type of the bulge is linked to intermediate-to-large scale environments.
    \item On scales less than $\sim 0.1h^{-1}$Mpc, the galaxies with a classic bulge present similar clustering properties and neighbour counts to control galaxies of similar mass, color and concentration, and this is true for all the scales probed and at all the masses except the lowest mass bin of $10^{9.5}<M_{stars}<10^{10}M_\odot$. In the lowest mass bin, galaxies of classic bulges appear to have more neighbours than the control galaxies, indicating that the presence of a classic bulge in low-mass galaxies is linked to galaxy-galaxy interactions or mergers.
    \item Galaxies with a pseudo bulge have more close neighbours within $\sim0.1 h^{-1}$Mpc when compared to control galaxies of similar mass, color and concentration, with the average neighbour count being enhanced by a factor of 2-3 at the smallest scales. The enhancement is weakly anti-correlated with stellar mass, with the sample-to-control $N_{neighbour}$ ratio slightly decreasing with increasing mass.
    \item pure disc galaxies with no bulges also show enhanced close neighbour counts within $\sim0.1 h^{-1}Mpc$ compared to the control galaxies, but the enhancement depends on stellar mass in a non-monotonic way, with the highest sample-to-control $N_{neighbour}$ ratio occurring at both high and low masses. As a consequence, galaxies at intermediate masses ($M_{stars}\sim 10^{9.5-10}M_\odot$) present the weakest signal. The $N_{neighbour}$ ratio increases to the levels of 5-6 at $M_{stars}>10^{10.5}M_\odot$ and $M_{stars}<10^{9.5}M_\odot$, an effect which is similar or even stronger than the previous measurement of the same quantify for the most strongly star-forming galaxies in the SDSS.
\end{itemize}

Note that in this work, we select only a very small fraction of galaxies with robust bulge type determinations based on the SDSS dr7 catalogue. Due to the limited numbers of galaxies selected, our matched samples are dominated by blue and low concentration galaxies (Fig.3). With future high spatial resolution observational data, and more accurate bulge-to-disc decomposition measurement that takes into account the components of nuclear and inner lenses/rings \citep{gao2017}, the clustering properties and other statistics of galactic bulges could be better derived, to set more constraints on the formation of galactic bulges. In addition, theoretical studies are also needed in order to have a full understanding of the formation process of bulges of both classical and pseudo types, as well as the physical relationship with pure-disc bulgeless galaxies. According to our study, these theoretical models must involve galaxy-galaxy interactions and other environment effects over a large range of spatial scales, as well as secular processes internal to galaxies.

\section*{Acknowledgements}

We thank the reviewer for the detailed and helpful comments. We thank Cheng Du, Shude Mao, Liang Gao and colleagues of the YIPA galaxy discussion group at NAOC for helpful discussions. LW acknowledges support from the NSFC grants program (No. 11573031), and the National Key Program for Science and Technology Research and Development (2017YFB0203300). CL is supported by National Key Basic Research Program of China (No. 2015CB857004) and NSFC (grant No. 11173045, 11233005, 11325314, 11320101002). CL, HJM and HYW are also supported by the National Key Basic Research and Development Program of China (No. 2018YFA0404502, No. 2018YFA0404503).

Funding for  the SDSS and SDSS-II  has been provided by  the Alfred P. Sloan Foundation, the Participating Institutions, the National Science Foundation, the  U.S.  Department of Energy,  the National Aeronautics and Space Administration, the  Japanese Monbukagakusho, the Max Planck Society,  and the Higher  Education Funding  Council for  England. The SDSS Web  Site is  http://www.sdss.org/.  The SDSS  is managed  by the Astrophysical    Research    Consortium    for    the    Participating Institutions. The  Participating Institutions are  the American Museum
of  Natural History,  Astrophysical Institute  Potsdam,  University of Basel,  University  of  Cambridge,  Case Western  Reserve  University, University of Chicago, Drexel  University, Fermilab, the Institute for Advanced   Study,  the  Japan   Participation  Group,   Johns  Hopkins University, the  Joint Institute  for Nuclear Astrophysics,  the Kavli Institute  for   Particle  Astrophysics  and   Cosmology,  the  Korean Scientist Group, the Chinese  Academy of Sciences (LAMOST), Los Alamos National  Laboratory, the  Max-Planck-Institute for  Astronomy (MPIA), the  Max-Planck-Institute  for Astrophysics  (MPA),  New Mexico  State University,   Ohio  State   University,   University  of   Pittsburgh, University  of  Portsmouth, Princeton  University,  the United  States Naval Observatory, and the University of Washington.

\bibliographystyle{mnras}
\bibliography{bulge}

\appendix
\section{uncertainty in the fitting of Sersic parameter}

\begin{figure}
\bc
\hspace{-0.4cm}
\resizebox{8cm}{!}{\includegraphics{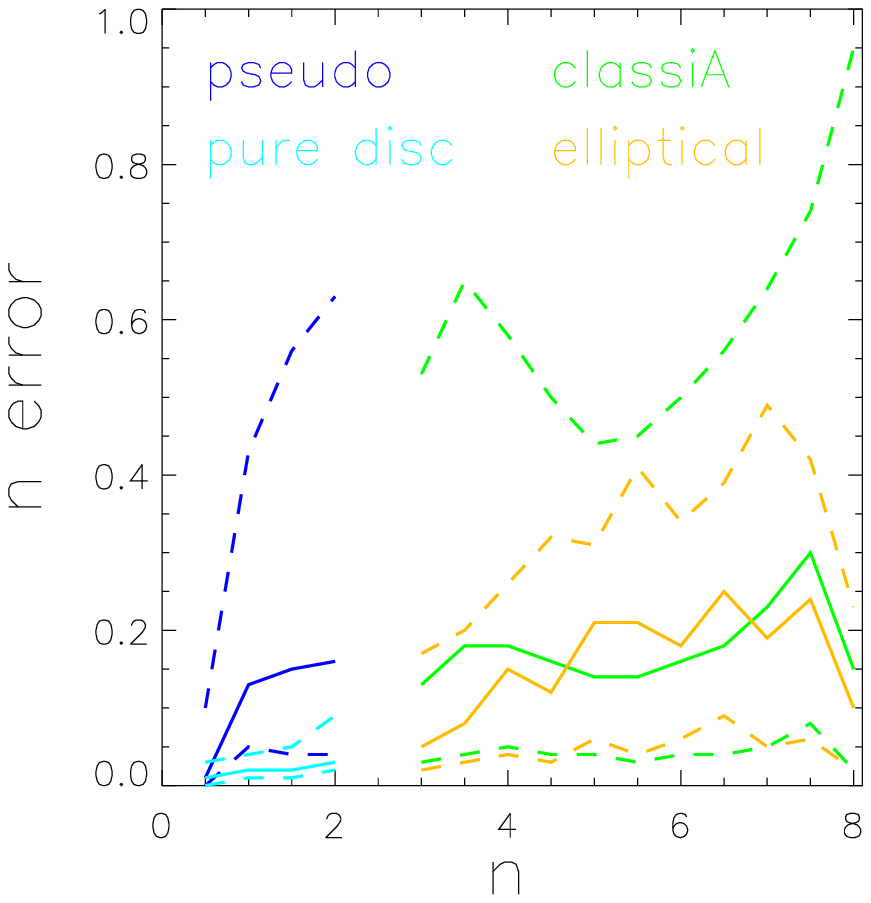}}\\
\caption{
The error of $n$ in our samples. Solid lines are the median error of n at given n value, while dashed lines include 68 per cent distribution of n error around the median.
}
\label{fig:nerror}
\ec
\end{figure}

In \citet{simard2011}, when fitting galaxies with a free $n_b$ bulge + disc model, and a pure Sersic $n_g$ model, the Sersic index $n$ is set to be within the range of $0.5<n<8$. In Figure.~\ref{fig:nerror}, we plot the distribution of error in $n$ in our samples, before excluding galaxies with $n \geq 7.95$. Classical B sample galaxies have $n_b=4$ with no error and are therefore not presented in the figure. Solid lines are the median value at given $n$, and dashed lines include 68 percentile distribution around the median. We have checked that when excluding galaxies with error in $n_b$ greater than the $+1 \sigma$ of error distribution (above the upper dashed lines) in our samples, and do the same matching of samples to make them have the same distribution of redshift, color and concentration, the resulting correlation function and neighbour count results remain similar. This means that these galaxies with large error in $n$ fitting do not affect our conclusion qualitatively. 

When galaxies host bars, the fitting bulge Sersic indices may be affected since the models adopted by \citet{simard2011} do not include the bar components. If the existence of bars significantly affect the fitting of Sersic indices, it should increase the resulting $n$ values and hence the fraction of classical bulge galaxies with bars should be larger than that of the pseudo bulge sample. To check the effect of bars on the determination of Sersic indices, we have cross-matched our samples with the morphological catalogue provided by \citet{nair2010}, which includes detailed visual classifications for 14,034 galaxies in SDSS. The matched number of galaxies in the pseudo bulge sample is 727, among which 26.5 per cent host bars. The matched number of galaxies in the classical bulge sample is 2702, and 21.9 per cent host bars. The percentage of barred system is a bit higher in the pseudo bulge sample. Therefore the existence of bars is not significantly increasing the $n$ values of the samples. 
Also note that the clustering as well as the neighbour counts of barred and unbarred galaxies of similar stellar mass is indistinguishable \citep{licheng2009,lin2014}, although bar growth and classical bulges may be related \citep{barway2016}. Therefore the existence of bars would not affect of the results of our selected galaxy samples.



\section{Kormendy diagram and velocity dispersion of our sample galaxies}

\citet{kormendy2016} provides the latest complete list of bulge classification criteria, which includes sSFR, velocity dispersion, bulge-to-total ratio, Sersic index, etc. Any one criterion has a non-zero probability of failure in determining the type of galactic bulges. Therefore the use of more classification criteria can give more reliable results. For example, \citet{vaghmare2015} combined the $n<2$ criterion with the offset in the Kormendy diagram \citep{kormendy1977} to select pseudo bulges.  \citet{mishra2017a} added one more criterion to select pseudo bulges to have velocity dispersion less than $130 km/s$ \citep{fisher2016}.

We should note, however, while the significant offset in the Kormendy diagram is useful to define typical pseudo bulges, identifying classical bulges using this relation is not robust \citep{fisher2016}, since some pseudo bulges also fit the Kormendy relation \citep{kormendy2016}. Besides, Kormendy relation can not be used to separate pseudo bulges from classical bulges in dwarf galaxies, since both types of bulge depart from this relation at low-luminosities \citep{graham2016}. On the other hand, including more criteria means including less number of galaxies in the selected samples, and would make the resulting galaxy samples without enough numbers to do statistics such as correlation functions and neighbour counts. Therefore, 
in this work, to select pseudo bulge and classical bulge galaxies, we use the Sersic index to define different bulge types. This is based on the fact that most pseudo bulges have Sersic index $n<2$, while most classical bulges have $n \geq 2$ \citep{kormendy2016,fisher2008,kormendy2004}. 

\begin{figure}
\bc
\hspace{-0.4cm}
\resizebox{8cm}{!}{\includegraphics{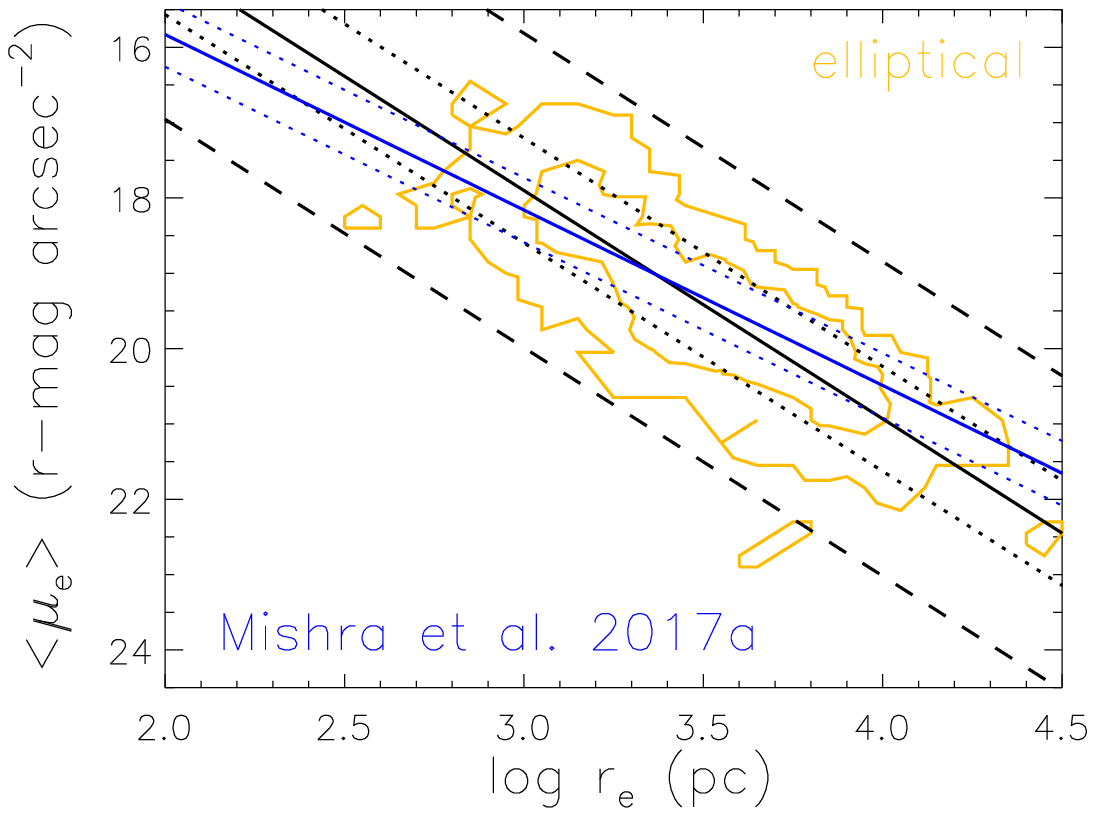}}\\
\caption{
r-band Kormendy diagram of our selected elliptical sample, as presented in section 2 and Table 1. Two contours include 68 and 95 percentile of galaxies in the sample. Black solid line is the best-fitting relation for our elliptical sample, while black dotted and dashed lines indicate 1 and 3 rms scatters respectively. As reference, blue solid line shows the best-fitting line for ellipticals using r-band data as presented by \citet{mishra2017a}, with blue dotted lines showing the range of their rms scatter.
}
\label{fig:elliptical}
\ec
\end{figure}

Nevertheless, to check whether using the Sersic index alone causes significant sample contamination and may be biasing our results, we analyse the Kormendy diagram of our selected samples. In Fig.~\ref{fig:elliptical}, we plot the r-band surface brightness - effective radius relation for our selected elliptical sample. Black solid line is the best-fitting relation for our elliptical sample, while black dotted and dashed lines indicate 1 and 3 rms scatters respectively. The result is in general consistent with the work of \citet{mishra2017a}, but with a bit larger scatter. 

\begin{figure}
\bc
\hspace{-0.4cm}
\resizebox{8cm}{!}{\includegraphics{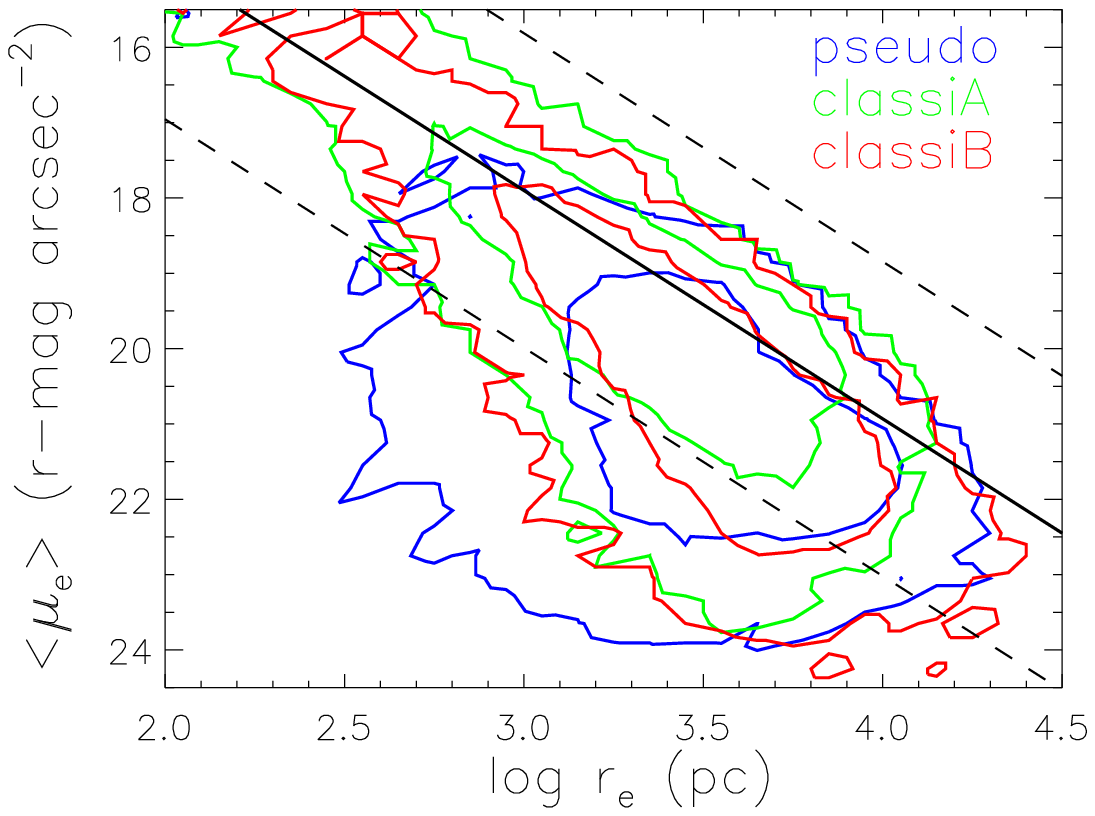}}\\
\caption{
r-band Kormendy diagram of our selected sample of pseudo bulge, classical bulge A and classical bulge B, as presented in section 2 and Table 1. For each sample, two contours include 68 and 95 percentile of galaxies in the sample. Black solid line is the best-fitting line for our elliptical sample, and dashed lines show the range of 3 times the rms scatter, which are the same as shown in Fig.~\ref{fig:elliptical}. 
}
\label{fig:kormendy}
\ec
\end{figure}

In Fig.~\ref{fig:kormendy}, we plot the position of galaxies on the Kormendy plane for our bulge samples. Most of the classical galaxies lie within the scatter of the Kormendy relation defined by our elliptical sample (dashed lines), with some exceptions lying below the relation. Pseudo bulges have a much larger fraction lying below the lower dashed line. In both \citet{vaghmare2015} and  \citet{mishra2017a}, pseudo bugles are required to lie below 3 times the rms of the best-fit line to the elliptical sample. Following their criterion, we select further our bulge samples according the Kormendy relation. Pseudo bulge galaxies with Kormendy criterion (pseudo-Kor) are required to be below the lower dashed line in Fig.~\ref{fig:kormendy}, and contain 30.0 per cent galaxies in our original pseudo bulge sample. Classical bulge galaxies with Kormendy criterion (classical-Kor) are required to be above the lower dashed line in Fig.~\ref{fig:kormendy}, and contain 90.0 per cent galaxies in our original classical bulge sample.

\begin{figure}
\bc
\hspace{-0.4cm}
\resizebox{8cm}{!}{\includegraphics{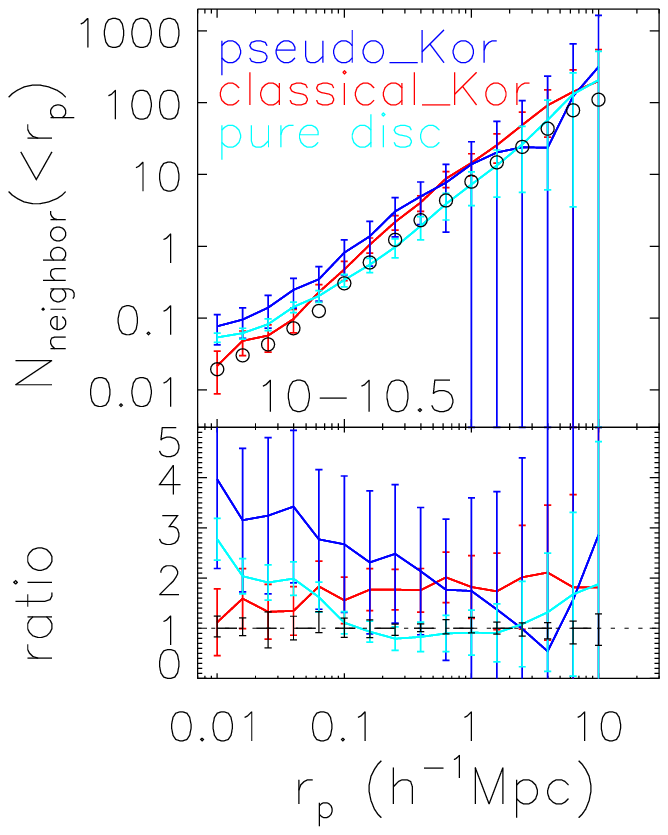}}\\
\caption{
Similar as Figure 5, but for classical bulge and pseudo bulge samples with additional constraint based on the Kormendy diagram (see text for the details). In each stellar mass bin, the samples are matched to have the same z, color and concentration distributions as the bulge samples and control samples shown in Figure 5. Due the limited numbers in the matched samples, only results at one stellar mass bin of $10^{10-10.5}M_{\odot}$ are shown here.
}
\label{fig:NC_kormendy}
\ec
\end{figure}

With the additional constraint based on the Kormendy diagram, we check the neighbour counts results of the pseudo-Kor and classical-Kor samples. To make a fair comparison, the galaxies in both the pseudo-Kor and classical-kor are matched with the samples shown in Fig.5 to have the same distributions in redshift, color and concentration in each stellar mass bin. The resulting numbers of galaxies in the two -Kor samples are small, which can give enough statistics only in the stellar mass bin of $10^{10-10.5}M_{\odot}$, and the result is shown in Fig.~\ref{fig:NC_kormendy}. Although the bootstrap errors of the two bulge samples are large, compared with the results in the panel of the same stellar mass in Fig.5, the general conclusion seems to still hold. Pseudo bulge galaxies have excess in neighbour counts on the small scales, while classical bulges have no excess on small scales. 

\begin{figure}
\bc
\hspace{-0.4cm}
\resizebox{8cm}{!}{\includegraphics{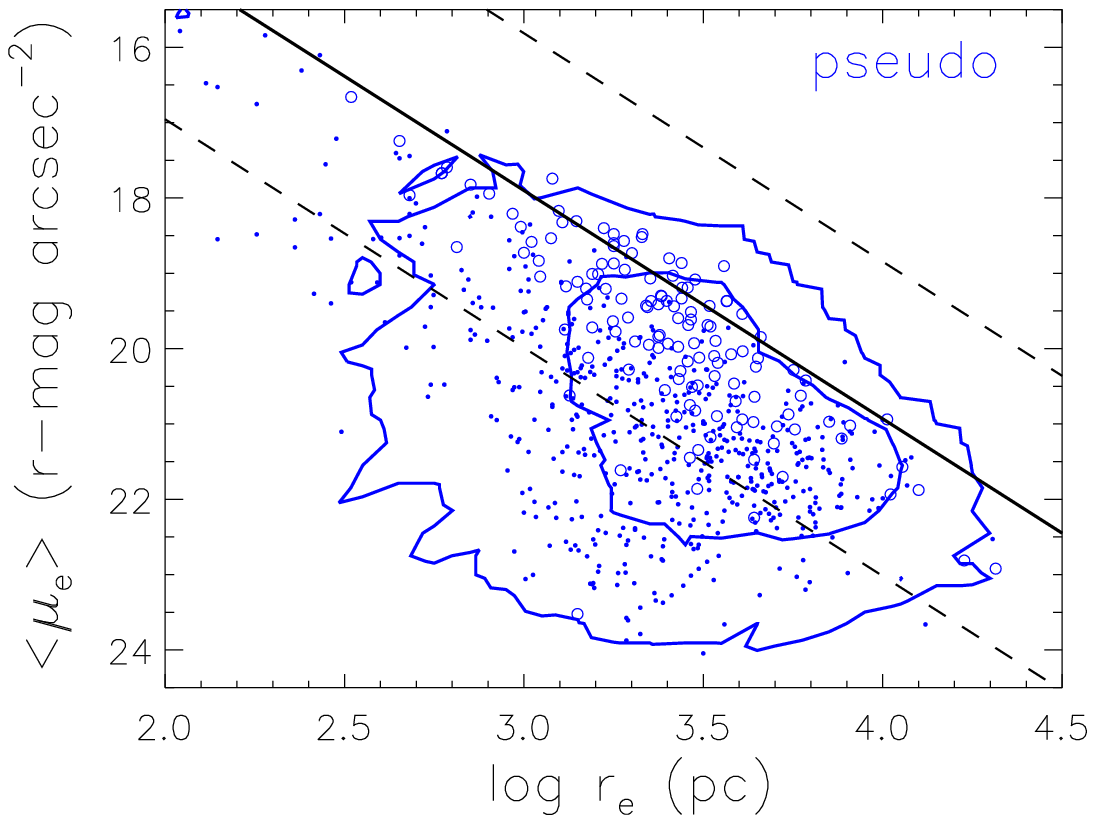}}\\
\caption{
r-band Kormendy diagram of our selected sample of pseudo bulge, same as in Figure B2. Over-plotted by dots and circles are pseudo bulge galaxies that have velocity dispersion information matched in the catalogue of \citet{nair2010}. Filled dots are galaxies with velocity dispersion less than 130km/s, while open circles are those with larger velocity dispersion.
}
\label{fig:velocity}
\ec
\end{figure}

To check the velocity dispersion of our sample galaxies, we have again matched our samples with the catalogue provided by \citet{nair2010}, which provide the kinematic information. In Fig.~\ref{fig:velocity}, we plot again the Kormendy diagram for our original pseudo bulge sample. The galaxies with velocity dispersion information (683 in number) are over-plotted, while solid dots represent galaxies with velocity dispersion less than $130 km/s$ (549 galaxies), and open circles are galaxies with larger velocity dispersion (134 galaxies). 227 galaxies with velocity dispersion information are below the lower dashed line, i.e., fulfilling the criterion based on the Kormendy relation, among which 97.4 per cent have velocity dispersion less than $130 km/s$. This test indicates that most of the pseudo bulge galaxies selected based on $n<2$ and have large offset on the Kormendy plane also match the velocity dispersion criterion. Nevertheless, as have discussed in the beginning of this section, these criteria are useful to select un-contaminated typical pseudo bulges, but do not include all pseudo bulges.

\bsp
\label{lastpage}


\end{document}